\documentclass[twocolumn,showpacs,prl,superscriptaddress]{revtex4}

\usepackage{graphicx}% Include figure files
\usepackage{dcolumn}% Align table columns on decimal point
\usepackage{bm}% bold math

\begin{document}

\preprint{APS/123-QED}

%
%
%
% symbolic definitions
%
\def\bfr{{\bf r}}
\def\bfk{{\bf k}}
\def\Shat{{\hat S}}
\def\calp{{\cal P}}
\def\calF{{\cal F}}
\def\qexp{{a^\prime}}
\def\DH{{\rm{Debye-H{\"u}ckel }}}
\def\CSM{{Departments of Chemistry and Chemical Engineering, Colorado School of Mines, Golden, CO 80401}}

%
%
%\voffset .75truein
%
% TITLE PAGE
\title{Invariance of density correlations with charge density in
polyelectrolyte solutions}
\author{James P. Donley}
\email{james.p.donley@boeing.com}
\affiliation{The Boeing Company, Huntington Beach, CA 92647}
\author{David R. Heine}
\altaffiliation[Present Address: ]{Sandia National Laboratories, Albuquerque, NM 87185}
\affiliation{\CSM}
\author{David T. Wu}
\affiliation{\CSM}

\date{\today}

\begin{abstract}
We present a theory for the equilibrium structure of polyelectrolyte solutions.  The
main element
is a simple, new optimization scheme that allows theories such as the random phase
approximation (RPA) to handle the harsh repulsive forces present in such systems.
Comparison is made with data from recent neutron scattering experiments of randomly
charged, hydrophilic polymers in salt-free, semi-dilute solution at various charge
densities.  Models with varying degrees of realism are examined.  The usual 
explanation of the invariance observed at high charge density has been counterion
condensation.
However, when polymer-polymer correlations are treated properly, we find that 
modeling polymer-counterion correlations at the level of \DH theory is sufficient.
\end{abstract}

\pacs{61.20.Gy,61.25.Hq,82.35.Rs,87.15.-v}% PACS, the Physics and Astronomy
\maketitle

% TEXT:
Polyelectrolytes are polymers with ionizable groups that dissociate in polar
solvent leaving ions bound to the chain and counterions free in solution.
The study of polyelectrolytes has traditionally been driven by their omnipresence in
living things \cite{moleccellbook}, but there has been rapidly growing 
use of and interest in synthetic ones for technological applications, typically at
the nano-scale level \cite{applications}.
The theoretical study of them is one of the most challenging in the field of 
complex fluids, however, because of the strong repulsions and attractions present
in such systems \cite{forster95}.

In this Letter, we present a scheme that allows a potentially large
class of theories to deal properly with these strong
long-range interactions.  We demonstrate the utility
of this approach by comparing with some recent intriguing experiments of
Nishida, Kaji and Kanaya (NKK) \cite{nishida95}, and
Essafi, Lafuma and Williams(ELW) \cite{essafi99}.  Counterion
condensation \cite{oosawamanning} commonly is invoked as the cause of the invariance
at high charge density observed in these experiments.
While there is some ambiguity in the literature about what precisely is counterion
condensation, it is generally regarded that the mechanism is not captured
in any theory in which ion-counterion correlations are included at the level of
\DH theory \cite{oosawamanning,ccrefs}.   However, we show in the theory described below
that \DH modeling of
ion-counterion correlations can itself produce an invariance at high charge density.

There are few theories that can handle the harsh repulsions and attractions
in polyelectrolyte solutions \cite{deshkovski01}.  Further, there is currently
no theory that provides reliable predictions for liquid structure in the semi-dilute
regime---which is the
region of most experimental interest.  The theory of Laria, Wu and Chandler \cite{laria91}
has been applied successfully
to examine density correlations there as a function of number density \cite{shew96},
but performs poorly at large, experimentally relevant, interaction energies and
charge densities \cite{dhwunpub}.
The theory of Donley, Rajasekaran and Liu (DRL) seems to perform properly
at large interaction energies and charge densities, but presently only for purely
repulsive systems \cite{donley98}.
Despite its simple form, the random phase approximation (RPA)
can in some instances give reasonable predictions for the free
energy of polyelectrolyte solutions \cite{rpafreeenergy}.  Here, we assume that
theories such as the RPA are also adequate for density correlations given
a suitable reinterpretion of the interaction potential that appears in
them \cite{andersen71}.  Let
\begin{equation}
 S_{MkM'k'}(r) = \langle \bigl [{\hat \rho}_{Mk}({\bf r}) - \rho_{Mk}\bigr]
                         \bigl [{\hat \rho}_{M'k'}(0) - \rho_{M'k'}\bigr]\rangle
\label{a1}
\end{equation}
be the density-density correlation function between a $k$ type monomer (or site) on
a molecule of type $M$ and a $k'$ type monomer on a molecule of type $M'$ a
distance $r = \vert {\bf r}\vert$ apart.  Here,
${\hat \rho}_{Mk}({\bf r})$ is the microscopic density of $k$-type monomers on
molecules of type $M$ at position $\bf r$ with average value $\rho_{Mk}$. The
brackets denote a thermodynamic average.
The Fourier transform of $S_{MkM'k'}(r)$ is the partial structure factor and links the
theory to scattering experiments.  The correlation function can be separated into
 $\it intra-$ and $\it inter-$molecular pieces,
$\Omega_{Mkk'}(r)$ and $H_{MkM'k'}(r)$:
\begin{equation}
S_{MkM'k'}(r) = \Omega_{Mkk'}(r)\delta_{MM'} + H_{MkM'k'}(r),
\label{a2}
\end{equation}
where $\delta_{MM'}$ is the Kronecker delta.  In this work, the molecular
structure function $\Omega_{Mkk'}(r)$ is assumed known.  The intermolecular correlation
function $H_{MkM'k'}(r) = \rho_{Mk}\rho_{M'k'}\bigl [ g_{MkM'k'}(r) - 1\bigr ]$,
where $g_{MkM'k'}(r)$ is the radial distribution function.  Thus, to compare with
experiment we need a form for this function in terms of the potentials and
molecular structures.  Consider for simplicity a system where all the interactions
can be decomposed as pair potentials $u_{MkM'k'}(r)$ between sites.  Here,
we take these potentials to be hard-core for distances $r$ less than some
range $\sigma_{MkM'k'}$, and Coulombic outside.  Consider also
a class of theories for which $g_{MkM'k'}(r)$ is a function only of the interactions
$\{u\}$ and the total molecular structure functions $\{\Omega\}$.  Theories in this
class include the RPA and possible higher loop improvements, and the
polymer version \cite{melenkevitz97} of the theory of
Chandler, Silbey and Ladanyi \cite{chandler82}.  One failure of theories such as the RPA is
that they predict a negative $g_{MkM'k'}(r)$ for short distances if the
interaction $u_{MkM'k'}(r)$ is strongly repulsive even though the radial distribution
function is intrinsically non-negative.  This shortcoming is thought to be
one reason that RPA underestimates the strength of the dependence of $q_{max}$ with
$\rho_m$ in the semi-dilute regime.

For short-range repulsive, e.g., hard-core, potentials one remedy to this deficiency
was proposed by Andersen and Chandler(AC) \cite{andersen71}.  Their idea was to replace the
true potential $u$ by a pseudo or optimized one, ${\tilde u}_{MkM'k'}(r)$.  For distances
$r < \sigma_{MkM'k'}$, ${\tilde u}_{MkM'k'}(r)$ takes a value such that the hard-core
condition is satisfied, i.e., $g_{MkM'k'}(r) = 0$.  Outside the core, ${\tilde u}_{MkM'k'}(r)$ can be set equal to the original potential $u$ which is long-ranged and
presumably weak.  In this way, higher order diagrams neglected in the theory are summed
in an approximate manner to enforce the core condition.  When applied to the RPA,
this optimization scheme is a generalization of the MSA closure of the Orstein-Zernike
equation \cite{lebowitz66} and a closely related scheme has been shown to be diagramatically
proper for CSL
theory \cite{melenkevitz97}.  This optimized RPA or ORPA, usually with long-range potentials
absent, is more popularly known as RISM for small molecules \cite{chandler82b} and PRISM for
polymers \cite{curro87}.

Unfortunately this AC optimization is not
very useful for polyelectrolytes.  One reason is that the contact energy between ion and counterion seems to
be less important than the interaction energy of the counterion with the whole chain.  Hence,
enforcing the hard-core behavior between opposite charges matters less than modeling
correctly the chain structure, e.g., determining the bond length $b$.  A more
important reason is that the repulsion between like-charged polymers is usually very
large.  This causes the radial distribution function to be effectively zero out to a
distance that scales with the \DH screening length for semi-dilute densities which is usually
much larger than the hard-core distance \cite{shew96,donley98,canessa91}.  Hence,
enforcing the core condition produces little improvement in the theory for
long polyelectrolytes.

An alternative
scheme is to optimize the ${\it range}$ of the pseudo-hard-core portion of $\tilde u$,
and not just its amplitude.  By this we mean that if
$g(r)$ is nearly zero out to some distance, then it makes no difference
whether this exclusion zone is caused by a hard-core potential or a Coulombic
one.  This effective hard-core diameter $\sigma^{eff}_{MkM'k'}$ is determined by
requiring that $g_{MkM'k'}(r)$ be non-negative everywhere (not just zero inside the core).
The closure to the theory then is:
\begin{equation}
	\begin{array}{ccc}
	g_{MkM'k'}(r) =& 0, & r < \sigma_{MkM'k'}^{eff},\\
	{\tilde u}_{MkM'k'}(r) =& u_{MkM'k'}(r), & r > \sigma_{MkM'k'}^{eff},
	\end{array}
\label{a3}
\end{equation}
and the range $\sigma_{MkM'k'}^{eff}$ of the hard-core interaction is chosen
to have the smallest value such that $g_{MkM'k'}(r) > 0$ for $r>\sigma_{MkM'k'}^{eff}$
subject to the constraint that $\sigma_{MkM'k'}^{eff} \geq \sigma_{MkM'k'}$.
With guesses for the $\sigma^{eff}_{MkM'k'}$'s, one can solve the RPA (or similar
theory) numerically in a manner very similar to what is done in PRISM.  One then
adjusts the $\sigma_{MkM'k'}^{eff}$ by iteration.
Eq. (\ref{a3}) is the primary result of this paper.

For simplicity, we show results for the range optimized version of RPA, denoted as RO-RPA.
We compare with scattering data of NKK and ELW.
In these experiments, the molecules were hydrophilic, sulfonated vinyl polymers
that were linear and randomly charged, with monovalent ions and counterions.
Small angle X-ray (SAXS) and neutron (SANS)
scattering methods were used to examine the equlibrium structure of polyelectrolytes and
counterions in their salt-free solution, water.
Previous experiments on
polyelectrolytes have examined how density correlations in the liquid change with
polymer monomer number density $\rho_m$.  Here, though, $\rho_m$ was held constant
at a semi-dilute or near semi-dilute density and the per chain average fraction of
charged monomers $f$ was varied.  Batches of polymers with different values of $f$ were
created and the scattered intensity $I(q)$ as a function of wavevector $q$ was
measured for each.
NKK found that at small $f$, the peak position of $I(q)$, $q_{max}$, obeyed a power law, but
at larger $f \sim 0.4$ appeared to reach an asymptote.  ELW extended these
measurements to higher $f$ and found that $q_{max}$ was effectively constant for
$f > 0.4$.  SANS measurements
also allowed them to extract the polymer monomer-monomer structure factor $\Shat_{mm}(q)$.
They found that at least for the range $0.55 < f < 0.81$, $\Shat_{mm}(q)$ was
invariant for the wavevectors measured.
A picture that ELW offer is that as one increases
$f$ above $0.4$, the system acts as if the charges on the chain are renormalized
such that the effective $f$ is constant.  Counterion condensation is usually 
given as the explanation.

The neutron scattering data of ELW on poly-AMAMPS is sufficiently well
characterized to allow quantitative comparison {\it via} $\Shat_{mm}(q)$ \cite{essafi99}.
We show results for $\Shat_{mm}(q)$ for three models,
the ``minimal", ``primitive" and ``two-state".  In all models the polymer chains are linear
and rod-like.  The chains are uniformly charged
and all monomers are identical so that the monomer valency $Z_m=f$, which
can be varied continuously from $0$ to $1$.
The strength of the system 
interactions is characterized partly by the ratio of the Bjerrum length
 $l_B = e^2/(\epsilon k_BT)$ to the polymer bond length $b$.
Here, $e$, $\epsilon$ and
$k_BT$ are the electron charge, solvent (water) dielectric constant and thermal
energy, respectively.
In the models here, the effect of the salt-free solvent enters only
through $\epsilon$.
Both monomer and counterion are spherical and we set their diameters, $\sigma$,
equal to each other and
equal to the bond length $b$.  For ELW, $\l_B\approx 7.1\AA$, $b\approx 2.5\AA$
and $\rho_m\sigma^3\approx 3\times 10^{-3}$.  The counterion density $\rho_c$ was
determined by charge neutrality, $Z_m\rho_m + Z_c\rho_c = 0$ and counterion valency
$Z_c = -1$.  The chain length N was set to 500.
In the minimal model the system consists only of linear polymers.
Monomers on the same chain or different chains interact with a potential
which is hard-core for $r<\sigma$.  Outside the core the potential has a screened,
\DH form: $u_{mm}(r) = {Z_m^2 l_B\over r} e^{-\kappa r}$,
where the inverse screening length $\kappa = \sqrt{4\pi Z_c^2 l_B \rho_c}$.
In the primitive model, the system consists of linear polymers and counterions.
Polymer monomers and counterions interact with a potential which is also hard-core
for $r<\sigma$.  Outside the core the potential has a bare Coulomb form:
$u_{ij}(r) = {Z_i Z_j l_B\over r}$,
where $i$ or $j$ denote monomers or counterions.

\begin{figure}
\includegraphics{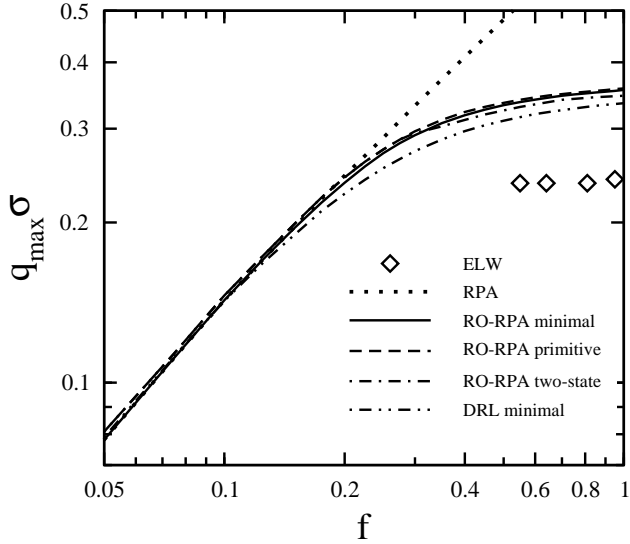}
\caption{\label{fig1} Scaled peak wavevector $q_{max}\sigma$ of the
monomer-monomer partial structure factor $\Shat_{mm}(q)$ as a function of
the average fraction of charged monomers per chain $f$ for $l_B/b=2.85$
and $\rho_m\sigma^3=3\times 10^{-3}$.  The meaning of the curves and symbols is shown
in the figure legend.
}
\end{figure}

Comparison of the predictions of RO-RPA in the primitive model with
simulation data \cite{stevens95} shows that RO-RPA
tends to underestimate the strength of polymer-counterion correlations \cite{dhwunpub}.
The two-state model is a common way to improve these
correlations \cite{oosawamanning,rpafreeenergy}.   In this model 
counterions are divided into two species, free and condensed.  In one version of
the model, free ones are counterions as in the primitive model,
and condensed ones are bound to the surface of the chain,
but able to translate along its length.  Each chain has the same number of condensed
counterions, $N_c$.  An average $N_c$ is determined by constructing a free energy
for the system and then minimizing it with respect to $N_c$.
 For a given $N_c$ we computed the
intramolecular structure functions $\Omega_{Mkk'}(r)$ for the polymer-condensed counterion
``molecule" by a single chain Monte Carlo simulation.  Here,
the intramolecular effective interaction between chain monomers and condensed counterions
was taken to have a screened \DH form similar to above, but the inverse screening
length $\kappa =
\sqrt{4\pi l_B(Z_m^2 \rho_m + Z_c^2\rho_c)}$ was due to both ions and counterions.
 We calculated the free energy using
the ``charging" formula \cite{chandler82b}, assuming that the molecular structure was
constant during the charging.  This constancy is not correct obviously; however, the
goal here is not a value for the free energy, but the position of its minimum with
respect to $N_c$.  We expect for the minimum that this approximation is a
reasonable one.

\begin{figure}
\includegraphics{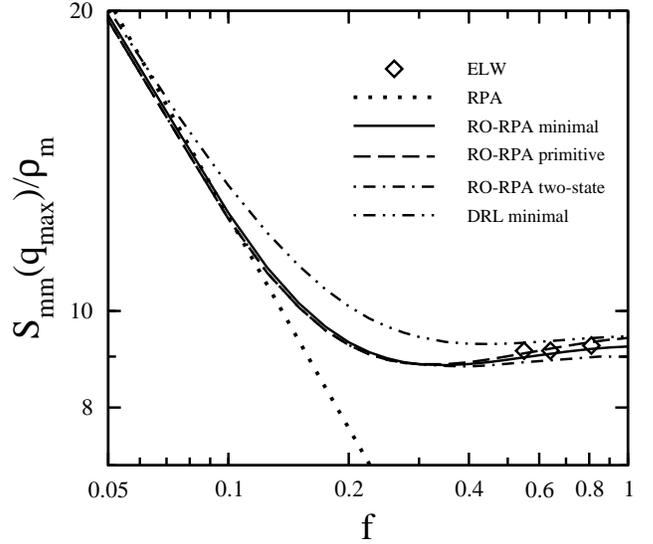}
\caption{\label{fig2} Scaled peak height of the monomer-monomer partial
structure factor $\Shat_{mm}(q_{max})/\rho_m$ as a function of 
the average fraction of charged monomers per chain $f$.  The conditions are the
same as in Figure 1.  The meaning of the curves and
symbols is shown in the figure legend.   
}
\end{figure}

Figures 1 and 2 show results for the peak position $q_{max}$ and
 peak height $\Shat_{mm}(q_{max})$ of the monomer-monomer partial structure factor.
The solid, dashed and dashed-dotted lines correspond to results
of RO-RPA using the minimal, primitive and two-state models,
respectively.
Results from the RPA are also shown as dotted lines.  Diamonds are data of ELW.
At small $f$, both quantities seem to obey
power laws: $q_{max} \sim f^\alpha$ and $\Shat_{mm}(q_{max}) \sim f^{-\eta}$.  All
RO-RPA models
predict $\alpha \approx 2/3$ and $\eta\approx 0.9$. 
For flexible hydrophilic polyelectrolytes $\alpha$ is expected to be
$1/3$ \cite{nishida95,pfeuty78}.  The difference is due probably to our modeling
the chains as rods while they are predicted to be rod-like only for a restricted
range of lengthscales \cite{pfeuty78}.
We know of no data for $\eta$ for polyelectrolytes.
At large $f \geq 0.4$, it is clear that range optimization 
completely changes the character of the RPA theory.  In contrast to the RPA,
RO-RPA in the minimal model predicts that both quantities become pretty much
constant, in agreement with NKK \cite{nishida95} and ELW \cite{essafi99}.
Note that adding explicit counterions and condensed counterions produces
only moderate improvement in the theoretical trends \cite{comment1}.
As a check on the validity of the RO-RPA theory, we also show results
(dot-dot-dashed lines) of the DRL theory \cite{donley98} in the minimal
model.
DRL is an approximation to the ``two-chain" equation for
$g(r)$ \cite{laria91,donley94}.  As can be seen, this theory also produces
an invariance at large $f$ and agrees almost
quantitatively with RO-RPA at all values of $f$.
Agreement between RO-RPA and experiment for $\Shat_{mm}(q_{max})$  is very good
with the former predicting values less than $5\%$ lower than the latter.
Agreement for $q_{max}$ is less satisfactory with
RO-RPA predicting values about $35\%$ higher than experiment.  The most probable
cause of this discrepancy is our neglect of chain flexibility since the \DH
screening length $\xi\sim 5\sigma$ at large $f$ implying a flexible chain on scales
larger than that.  If the chains were partially collapsed, then the repulsion and
average distance between monomers on different chains would be larger.
Thus $q_{max}$ would decrease and $\Shat_{mm}(q_{max})$ would increase.
On the other hand, if the solvent---which is ``good" for
poly-AMAMPS---were included explicitly then $\Shat_{mm}(q_{max})$ would decrease.
An explanation for this invariance with $f$ seen by the theory is due to
screening:  as $f$ increases,
the repulsion between polymer chains increases; on the other hand, charge
neutrality forces more and more counterions into the solution which increases
the screening between polymers.  At small $f$ polymer and counterion correlations
are weak, but at large $f$ they increase such that one reaches a balance between
polymer-polymer repulsion and counterion screening.

In conclusion, we have presented a new optimization scheme that appears to
improve substantially the predictive power of the RPA theory for the
structure of polyelectrolyte solutions.
We showed that this range-optimized RPA in the minimal model yields all
trends exhibited in the experimental data of ELW
at large $f$ and is in moderate agreement with that data for
the peak position and height.
The crudeness of our chain model, however, precludes us from drawing any firm
conclusions about the origins of the effects seen in ELW.  On the other hand,
since the minimal model $\it does$ predict an invariance and the comparison
is reasonable enough, it is apparent that a number of questions need to
be answered before the cause(s) of this phenomenon is determined.
But it is interesting that a model of screened Coulomb chains can produce effects
that would normally be associated with more complex mechanisms such as counterion
condensation.

\begin{acknowledgments}
We thank  John G. Curro, Andrea J. Liu, John D. McCoy,
Monica Olvera de la Cruz and Craig E. Pryor for helpful conversations and 
correspondence.
\end{acknowledgments}

%end text
%
%----------------------------------------------------------------------

%----------------------------------------------------------------------%

%\begin{thebibliography}

%\end{thebibliography}
%
%
%\begin{figure}
%\label{fig1}
%\caption{
%}
%\end{figure}
%
%

%--------------------------------------------------------------------------

%

\begin{references}
\bibitem{comment1}Interestingly, we find that the number of condensed counterions
per monomer $f_{cc} = 0$ for $f<0.3$, but rises to $0.45$
for $f=1$.
\end{references}
\end{document}